\documentclass[sn-mathphys,iicol]{sn-jnl}

\usepackage{physics}
\usepackage{cleveref}

\jyear{2022}%

\theoremstyle{thmstyleone}%
\theoremstyle{thmstyletwo}%

\theoremstyle{thmstylethree}%

\usepackage[numbers]{natbib}

\usepackage{accents}
\newlength{\dhatheight}
\newcommand{\doublehat}[1]{%
    \settoheight{\dhatheight}{\ensuremath{\hat{#1}}}%
    \addtolength{\dhatheight}{-0.35ex}%
    \hat{\vphantom{\rule{1pt}{\dhatheight}}%
    \smash{\hat{#1}}}}

\begin{document}

\title[\space]{Optimisation-free Density Estimation and Classification with Quantum Circuits}


\author*[1]{\fnm{Vladimir} \sur{Vargas-Calderón}}\email{vvargasc@unal.edu.co}

\author[2]{\fnm{Fabio A.} \sur{González}}

\author[1]{\fnm{Herbert} \sur{Vinck-Posada}}
\affil[1]{\orgdiv{Grupo de Superconductividad y Nanotecnología, Departamento de Física}, \orgname{Universidad Nacional de Colombia}, \orgaddress{\postcode{111321}, \city{Bogotá}, \country{Colombia}}}
\affil[2]{\orgdiv{MindLab Research Group, Departamento de Ingeniería de Sistemas e Industrial}, \orgname{Universidad Nacional de Colombia}, \orgaddress{\postcode{111321}, \city{Bogotá}, \country{Colombia}}}


\abstract{We demonstrate the implementation of a novel machine learning framework for probability density estimation and classification using quantum circuits. The framework maps a training data set or a single data sample to the quantum state of a physical system through quantum feature maps. The quantum state of the arbitrarily large training data set summarises its probability distribution in a finite-dimensional quantum wave function. By projecting the quantum state of a new data sample onto the quantum state of the training data set, one can derive statistics to classify or estimate the density of the new data sample. Remarkably, the implementation of our framework on a real quantum device does not require any optimisation of quantum circuit parameters. Nonetheless, we discuss a variational quantum circuit approach that could leverage quantum advantage for our framework.}




\maketitle

\section{Introduction}\label{sec1}

Quantum machine learning (QML) is regarded as an early application of noisy intermediate-scale quantum computing that could leverage quantum advantage~\citep{bharti2022noisy,huang2021information,huang2021power,huang2021quantum}.
During the last few years, there have been several proposals to perform different supervised and unsupervised QML tasks~\citep{rebentrost2014quantum,schuld2015introduction,Dunjko2018machinelearning,biamonte2017quantum,bharti2022noisy}. 
Most of the QML literature focuses on gradient-based algorithms that rely on hybrid approaches whereby a classical computer is used to update variational parameters of quantum circuits to minimise a given cost function~\citep{Benedetti2019parameterized,caro2021generalization}.
However, gradient descent--applied to quantum circuits--is known for scaling poorly with the number of qubits, as the probability of the gradients being non-zero is exponentially small as a function of the number of qubits~\citep{mcclean2018barren}.
This phenomenon is commonly addressed as the barren plateau problem and jeopardises the practical achievement of quantum advantage.
For this reason, there has also been a general interest of using gradient-free techniques to train variational quantum circuits~\citep{franken2020gradient,benedetti2019generative,peruzzo2014variational,khatri2019quantum,leyton2021robust} (despite some controversy~\citep{Arrasmith2021effectofbarren,marrero2020entanglement,cerezo2021cost}), as well as coming up with quantum-inspired gradient-free machine learning methods that can run both on classical and quantum computers~\citep{gonzalez2021classification,gonzalez2021learning,sergioli2019binary}.

In this work, we report the implementation of an optimisation-free framework~\citep{gonzalez2021classification,gonzalez2021learning}, based on a kernel approximation strategy~\citep{mengoni209kernel,schuld2019qml,havlivcek2019supervised,blank2020quantum}, on real quantum devices for density estimation and classification\footnote{We also release a library that is used to perform local and remote (on IBM quantum computers) runs of quantum circuits for density estimation and classification: \url{https://gitlab.com/ml-physics-unal/qcm}}.
This framework can be used for supervised and unsupervised machine learning tasks, which we exemplify through density estimation and classification.
Its main feature is that a data set of arbitrarily many samples can be compressed into a quantum state of a fixed number of qubits.
Once this quantum state is prepared, it is projected onto a quantum state of a sample that is to be classified or whose density is to be estimated.
The latter quantum state is built using a quantum feature map encoding.
Therefore, classification or density estimation (unlike many quantum kernel methods) can be achieved by just a single estimation of a quantum state overlap between a quantum state that encodes an arbitrarily large data set and the corresponding quantum state of the sample of interest. 

This paper is organised as follows. \Cref{sec:method} explains the optimisation-free framework for density estimation and classification. \Cref{sec:circuit} shows how to perform these tasks on a quantum computer. Then, \cref{sec:results} presents results for a couple of experiments carried out on real quantum devices. After that, \cref{sec:discussion} discusses the results in light of future challenges. Finally, \cref{sec:conclusions} concludes.

\section{Optimisation-free Density Estimation and Classification}\label{sec:method}

In this section, we outline the algorithms for density estimation and classification based on quantum measurements performed on general physical systems, which can be efficiently simulated in classical computers~\citep{gonzalez2021classification,gonzalez2021learning}.

The departure point for both algorithms is the availability of a quantum feature map (QFM) $\psi: \mathcal{X} \to \mathcal{H}_\mathcal{X}$, where $\mathcal{X}$ is the space of classical data features, and $\mathcal{H}_\mathcal{X}$ is the Hilbert space of some physical system.
Thus, the QFM maps a data sample to the quantum state of a physical system, i.e. $\psi: \vb*{x}_i\mapsto \ket{\psi_i} = \psi(\vb*{x}_i)$, where $i$ indexes a set of data samples.

A quantum state for a data set of $N$ samples $\mathscr{D}=\{\vb*{x}_i\}_{i=1,\ldots,N}$ can be built through
\begin{align}
    \ket{\Psi} = \mathcal{N}^{-1}\sum_{i=1}^N \ket{\psi_i}, \label{eq:datasetState}
\end{align}
where $\mathcal{N}$ is a normalisation constant. \Cref{eq:datasetState} shows that the data set state is a superposition of the states corresponding to each sample.

\subsection{Density Estimation}

Density estimation can be seen as the question: how likely is it that a point $\vb*{x}_\star\in\mathcal{X}$ is sampled from a distribution from which a data set $\mathscr{D}$ has already been sampled?
The quantum state in~\cref{eq:datasetState} encodes an estimation of the underlying probability distribution of the training data set.
The structure of the estimated probability distribution is given by the QFM.
This view is rather useful in quantum mechanics, as the state $\ket{\Psi}$ is naturally related to a probability distribution.
We can evaluate this estimation through the usual Born rule, i.e. at a point $\vb*{x}_\star$, the estimated probability density is
\begin{align}
    \hat{f}(\vb*{x}_\star) = \abs{\bra{\Psi}\psi(\vb*{x}_\star)}^2 =  \abs{\braket{\Psi}{\psi_\star}}^2. \label{eq:ProbDensEst}
\end{align}

\Cref{eq:ProbDensEst} exploits the especial relation between probability and geometry in quantum mechanics, whereby purely geometrical operations result in probability estimations.
More explicitly, the prediction will be $\abs{\mathcal{N}^{-1}\sum_{i=1}^N\braket{\psi_i}{\psi_\star}}^2$, where--due to the superposition property shown in~\cref{eq:datasetState}--the argument of the square modulus resembles the Parzen-Rosenblatt estimator~\citep{parzen1962,rosenblatt1956} if $\braket{\psi_i}{\psi_\star} = k(\vb*{x}_i - \vb*{x}_\star)$, for some kernel function $k$.
Examples of such kernel functions will be given in~\cref{sec:results}.

\subsection{Classification}\label{sec:theoryClassification}

To incorporate a class $y_i\in\mathcal{Y}$ for each sample $\vb*{x}_i\in\mathcal{X}$, we consider another QFM $\phi:\mathcal{Y}\to\mathcal{H}_\mathcal{Y}$, where $\mathcal{Y}=\{1,\ldots,K\}$ is a discrete set of $K$ elements or classes, and $\mathcal{H}_\mathcal{Y}$ is the Hilbert space of some physical system. Therefore, a labelled data set $\mathscr{C}=\{(\vb*{x}_i, y_i)\}_{i=1,\ldots,N}$ can be mapped to a quantum state via
\begin{align}
    \ket{\Psi} = \mathcal{N}^{-1} \sum_{i=1}^N \ket{\psi_i}\otimes \ket{\phi_i},\label{eq:classificationDataset}
\end{align}
where $\ket{\phi_i} = \phi(y_i)$.

As explained in Ref.~\citep{gonzalez2021classification}, the classification of a new data point $\vb*{x}_\star$ consists of projecting the $\mathcal{X}$ part of the data set quantum state onto the corresponding new data point quantum state $\ket{\psi_\star}$. 
More formally, we represent the state of the new data point in the combined space $\mathcal{H}_\mathcal{X} \otimes \mathcal{H}_\mathcal{Y}$ as $\ketbra{\psi_\star}\otimes \text{Id}_\mathcal{Y}$, where $\text{Id}_\mathcal{Y}$ represents the unknown state of the $\mathcal{Y}$ component. We project this state onto the data set state $\ketbra{\Psi}$.   After normalising, we trace out the degrees of freedom corresponding to the $\mathcal{X}$ part of the quantum system, leaving a reduced density matrix
\begin{align}
    \rho_\mathcal{Y}(\vb*{x}_\star) = \Tr_\mathcal{X}\left(\frac{\ketbra{\Psi}(\ketbra{\psi_\star}\otimes \text{Id}_\mathcal{Y})}{\Tr[\ketbra{\Psi}(\ketbra{\psi_\star}\otimes \text{Id}_\mathcal{Y})]}\right)
    \label{eq:partialTraceState}
\end{align}
from which we can obtain the probability $P(k\vert\vb*{x}_\star) =\bra{\phi_k}\rho_\mathcal{Y}(\vb*{x}_\star)\ket{\phi_k}$ that $\vb*{x}_\star$ is of the class $k$.
Note that $\rho_\mathcal{Y}(\vb*{x}_\star)$ contains all the probabilities of $\vb*{x}_\star$ belonging to any class.
However, we can directly calculate the probability through yet another application of the Born rule $P(k\vert\vb*{x}_\star) = \abs{\bra{\Psi}(\ket{\psi_\star}\otimes\ket{\phi_k})}^2$.

\section{Circuit Implementation}\label{sec:circuit}

In \cref{sec:circuitDE,sec:CircuitClassification} we will show how density estimation and classification can be carried out when the QFMs map classical data onto the state of a multi-qubit system.
\Cref{sec:cicuitPrep} will discuss how the particular quantum circuit unitaries can be implemented to perform density estimation and classification.

\begin{figure}
    \centering
    \includegraphics[width=\columnwidth]{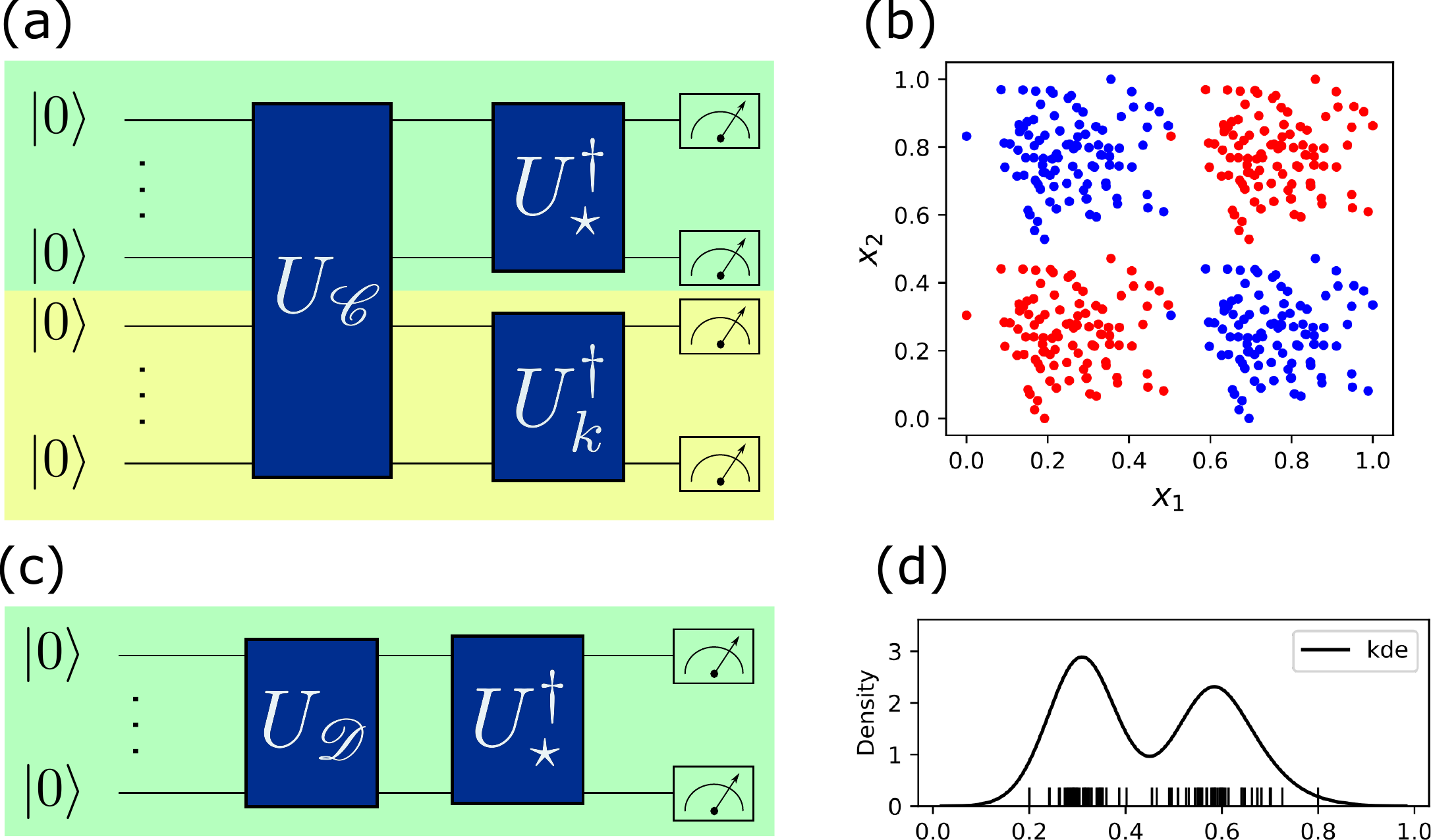}
    \caption{Quantum circuits and data sets for density estimation and classification. (a) is the circuit for classification of a point $\vb*{x}_\star$ given a training data set $\mathscr{C}$; the green part of the circuit corresponds to the data feature space $\mathcal{X}$ and the yellow one to the labels space $\mathcal{Y}$. (b) shows a toy data set used for classification, with two features $x_1$ and $x_2$ and a label shown as the red or blue colour. (c) is the circuit for estimating the probability density of a point $\vb*{x}_\star$ given a training data set $\mathscr{D}$. (d) shows a 1D toy data set used for density estimation and a kernel density estimation (KDE) fit.}
    \label{fig:circuits}
\end{figure}

\subsection{Density Estimation}\label{sec:circuitDE}

A general quantum circuit for probability density estimation is given in~\cref{fig:circuits}(c), where the probability density at a point $\vb*{x}_\star$ is computed using a training data set $\mathscr{D}$. 
Similarly as in classification, the quantum circuit can be seen in two ways in order to grasp which states are being prepared: from left to right, the unitary $U_\mathscr{D}$ prepares the data set quantum state $\ket{\Psi} = U_\mathscr{D}\ket{0}^{\otimes N_\mathcal{X}}$; and from right to left, the unitary $U_\star$ prepares the sample data point quantum state $\ket{\psi_\star} = U_\star\ket{0}^{\otimes N_\mathcal{X}}$.
Here, $N_\mathcal{X}$ is the number of qubits used to represent the data features.
Thus, the complete circuit prepares the state $U_\star^\dagger U_\mathscr{D}\ket{0}^{\otimes N_\mathcal{X}}$, whose projection onto $\ket{0}^{\otimes N_\mathcal{X}}$ gives the probability density at $\vb*{x}_\star$.

The latter procedure allows the direct estimation of the probability density as shown in~\cref{eq:ProbDensEst} by making $M$ measurements of the quantum circuit and by computing $\doublehat{f}(\vb*{x}_\star)=M_{\vb*{0}}/M$, where $M_{\vb*{0}}$ is the number of times that the $\vb*{0}$ bit string is measured. Explicitly, the complete protocol can be carried out as follows:
\begin{enumerate}
    \item Given a QFM $\psi$, compute $\ket{\psi_i} = \psi(\vb*{x}_i)$ for each data sample in $\mathscr{D}$.
    \item Compute the training data set using~\cref{eq:datasetState}.
    \item Use an arbitrary state preparation algorithm (see~\cref{sec:cicuitPrep} to get the circuit $U_\mathscr{D}$ that prepares the state in~\cref{eq:datasetState} on a quantum computer.
    \item Compute $\psi(\vb*{x}_\star)$ for a new data point $\vb*{x}_\star$.
    \item Use the arbitrary state preparation algorithm in step 3 to get the circuit $U_\star$ that prepares $\psi(\vb*{x}_\star)$.
    \item Run the circuit depicted in~\cref{fig:circuits}(c) $M$ times to estimate $M_{\vb*{0}}/M$ with whatever required precision you need.
\end{enumerate}

Note that once the state in~\cref{eq:datasetState} has been computed, there is no need to perform steps 1-3 to estimate the density of new data points.

\subsection{Classification}\label{sec:CircuitClassification}

A general quantum circuit for classification is depicted in~\cref{fig:circuits}(a), where the probability that a new data point $\vb*{x}_\star$ is of class $k$ is computed using a training labelled data set $\mathscr{C}$.
The quantum circuit can be seen as follows. From left to right, the unitary $U_\mathscr{C}$ prepares the quantum state of the data set $\mathscr{C}$: $\ket{\Psi} = U_\mathscr{C} \ket{0}^{\otimes N_\mathcal{X} + N_\mathcal{Y}}$, where $N_\mathcal{Y}$ is the number of qubits used to represent the data labels.
From right to left, the unitary $U_\star\otimes U_k$ prepares the quantum state of the new data point along the $k$-th class direction: $\ket{\psi_\star}\otimes\ket{\phi_k} = U_\star\otimes U_k\ket{0}^{\otimes N_\mathcal{X} + N_\mathcal{Y}}$. 
Therefore, the quantum circuit prepares a state $(U^\dagger_\star\otimes U^\dagger_k) U_\mathscr{C}\ket{0}^{\otimes N_\mathcal{X} + N_\mathcal{Y}}$ such that its projection onto $\ket{0}^{\otimes N_\mathcal{X} + N_\mathcal{Y}}$ gives the probability of $\vb*{x}_\star$ being classified in class $k$.

Thus, the classification probability can be estimated by sampling the quantum circuit $M$ times and counting the number of times $M_{\vb*{0}}$ that the $\vb*{0}$ bit string is measured.  
Then, the estimated probability is $\hat{P}(k\vert\vb*{x}_\star) = M_{\vb*{0}} / M$.

To summarise, a recipe similar to the one shown in~\cref{sec:circuitDE} can be followed to perform classification:
\begin{enumerate}
    \item Given a pair of QFMs $\psi$ and $\phi$ for data features and labels, compute $\psi(\vb*{x}_i)$ and $\phi(y_i)$ for every pair in the classification data set $\mathscr{C}$.
    \item Compute the training data set using \cref{eq:classificationDataset}.
    \item Use an arbitrary state preparation algorithm (see~\cref{sec:cicuitPrep}) to get the circuit $U_\mathscr{C}$ that prepares the state in \cref{eq:classificationDataset} on a quantum computer.
    \item Compute $\psi(\vb*{x}_\star)$ for a new data point $\vb*{x}_\star$.
    \item Use the arbitrary state preparation algorithm in step 3 to get the circuit $U_\star$ that prepares $\psi(\vb*{x}_\star)$.
    \item Run the circuit depicted in~\cref{fig:circuits}(a), where $U_k$ is not required if $\phi$ is the one-hot encoding, as explained in the appendix~\ref{sec:OHEQFM}. This circuit has to be run $M$ times to estimate $\hat{P}(k\vert\vb*{x}_\star)$ with a desired precision.
\end{enumerate}

As in~\cref{sec:circuitDE}, once the state in~\cref{eq:classificationDataset} has been computed, there is no need to perform steps 1--3 to classify a new data point.

\subsection{Multi-Qubit Quantum State Preparation}\label{sec:cicuitPrep}

The method that we have so far explored depends on the ability to compile the unitaries $U_X$ for $X=\mathscr{C},\mathscr{D}, \star, k$ into quantum circuits readable by current quantum computers. 
Most current quantum computers have primitive one- and two-qubit gates that allow universal quantum computation. 
Therefore, even though the general unitary $U_X$ is known, we need to decompose it into the primitive quantum gates of a quantum computer.

Several algorithms for arbitrary unitary decomposition have been suggested~\citep{barenco1995qrdecomposition,mottonen2004sincosDecomposition,krol2022efficient,li2013decomposition}. 
In this work, we use the algorithm proposed in Ref.~\citep{shende2006synthesis}, that offers a preparation of an $n$-qubit state using at most $2^{n+1}-2n$ CNOT gates.
This algorithm is implemented in the popular library for quantum computing Qiskit~\citep{Qiskit}, which we used to connect to publicly available quantum computers from IBM.

\begin{figure*}
    \centering
    \includegraphics[width=\textwidth]{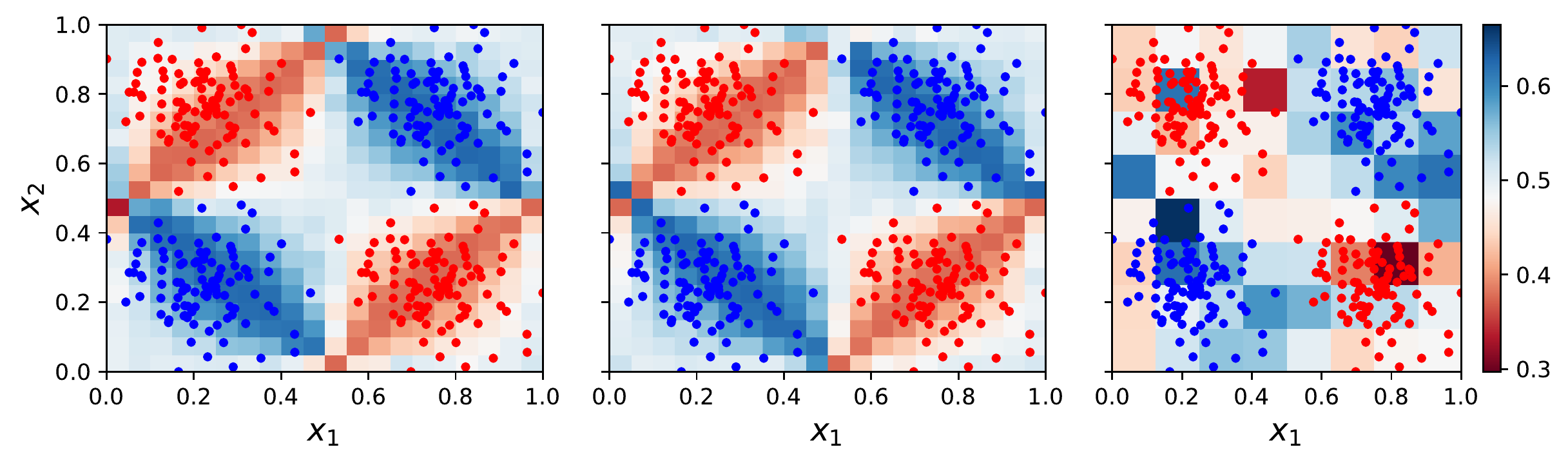}
    \caption{Predictions (background colour) of exact circuit simulation (left), noisy circuit simulation (middle, see main text for details on the noise model) and circuit on the IBM Bogotá quantum device (right) for a XOR data set (points, cf.~\cref{fig:circuits}(b)). 
    The colour indicates the probability that a point is classified in the blue class, as shown by the colour bar.
    The area under the the receiver operating characteristic curve was 99.93\%, 99.82\% and 95.83\% for the predictions of the exact simulation, noisy simulation, and real quantum device, respectively.}
    \label{fig:XORPreds}
\end{figure*}

Remarkably, recent work has produced new ways to prepare arbitrary quantum states using shallow quantum circuits~\citep{bausch2020fast}, by using additional ancillary qubits~\citep{araujo2021qsp,zhang2021lowdepth}, by training parametrised quantum circuits in the so-called quantum machine learning setup~\citep{schuld2019qml,Haug2020classifying,rakyta2022efficient}, or even by implementing tensor-network inspired gradient-free optimisation techniques~\citep{shirakawa2021automatic}.

Regardless of the quantum state preparation algorithm, our density estimation and classification framework retains the advantage of condensing the complete, arbitrarily large data set into a single quantum state of fixed size.


\section{Results}\label{sec:results}

The quantum circuit shown in \cref{fig:circuits}(a) was used to classify data in a XOR disposition, as shown in~\cref{fig:circuits}(c). 
Such toy data set is able to tell apart linear classifiers from non-linear classifiers. 
In our case, non-linearity is induced by the QFM. 
As an example, we consider the following QFM 
\begin{align}
\psi(x_1, x_2) = \bigotimes_{i=1}^2 (\sin\pi x_i\ket{0} + \cos\pi x_i\ket{1}).\label{eq:sincosQFM}
\end{align}
which ensures that the induced kernel $\abs{\psi^\dagger(\vb*{x})\psi(\vb*{x}^\prime)}^2$ is a pairwise cosine-like similarity measure $\cos\pi(x_1 - x^\prime_1)\cos\pi(x_2 - x^\prime_2)$. 
Regarding class labels, we selected the one-hot encoding as the QFM, such that red points are mapped to $\ket{0}$ and blue points are mapped to $\ket{1}$. Thus, a total of three qubits are used to perform the classification quantum circuit, with two qubits encoding the data features, and the remaining one encoding the class label.

\Cref{fig:XORPreds} shows three panels that display the probability that a point placed in $[0,1]\times[0, 1]$ is assigned to the red class or the blue class. 
The three panels correspond to a classical simulation of the classification quantum circuit on the left, a classical simulation of the corresponding noisy quantum circuit on the middle, and the classification carried out on the IBM Bogotá quantum device on the right.
The noise model for the quantum circuit, as modelled by IBM's Qiskit~\citep{Qiskit}, applies imperfect gates that have been fit to experimental measurements to a Krauss noise model~\citep{bogdanov2013}.
It is worth noting that the state of a noisy quantum circuit is never described by a state vector.
Instead, it is described by a density matrix.
However, this is not exactly how noise is modelled in Qiskit.
Imperfect gates are applied instead of the ideal ones with a previously measured probability distribution for selecting the gate to apply.
This way of applying noisy gates is analogous to the quantum trajectory approach~\citep{dalibard1992,molmer1993monte}, where the state vector of the quantum circuit is updated with the application of gates (called collapse operators in the open quantum system literature).
The true quantum state, described by a density matrix, can be recovered by averaging many realisations of the noisy quantum circuit, which is done by running the stochastic quantum circuit many times.
The noisy processes that are taken into account are single-qubit readout errors, reset errors, single-qubit Pauli- and $\sqrt{S}$-gate errors and, two-qubit C-NOT gate errors~\citep{chow2015}.
It is clear from the middle and right panels of~\cref{fig:XORPreds} that the used noise model is not able to simulate the real noisy quantum circuit, most likely because such a simplified noise model does not account for the complex dynamics that the quantum circuit undergoes as an open quantum system~\citep{berg2022probabilistic}.

\begin{figure*}
    \centering
    \includegraphics[width=\textwidth]{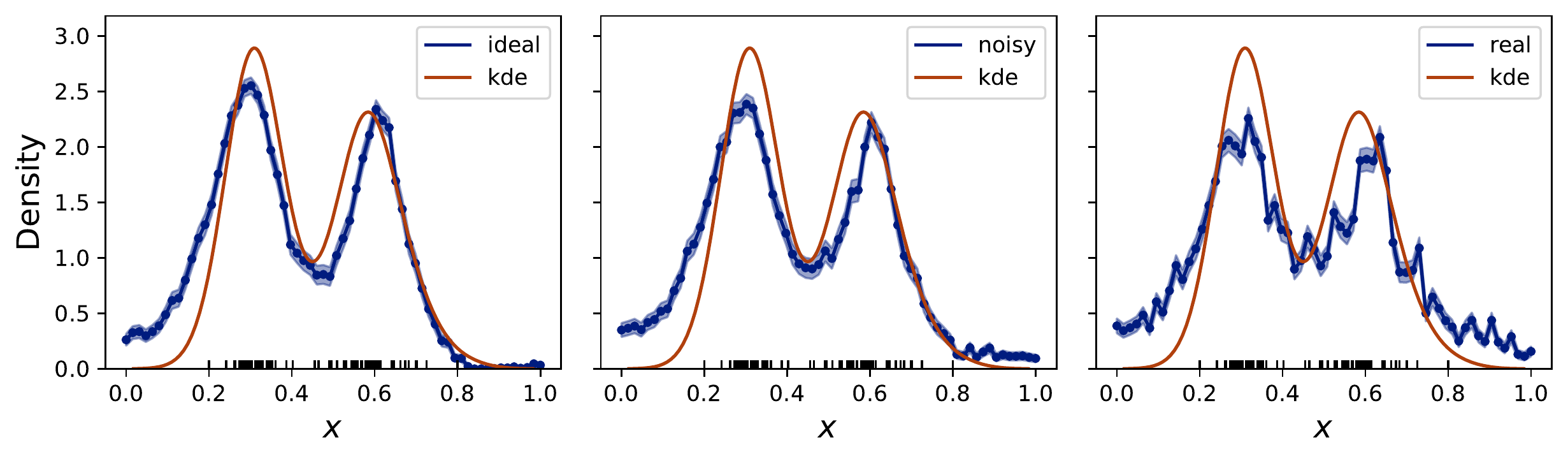}
    \caption{Density estimation (blue points) of bi-Gaussian-distributed data (cf.~\cref{fig:circuits}(d)) with exact circuit simulation (left), noisy circuit simulation (middle, see main text for details on the noise model) and run on the IBM Lima quantum computer. Orange lines are computed through regular Gaussian kernel density estimation. 1024 shots were used to estimate every point on a (simulated or real) quantum computer. Confidence intervals are computed with the asymptotic normal approximation of the Bernoulli distribution from which measurements are sampled.}
    \label{fig:GaussianPreds}
\end{figure*}

A more general QFM that is not as hand-tailored as the one introduced in~\cref{eq:sincosQFM} is the random Fourier features (RFFs) QFM that we proposed in Ref.~\citep{gonzalez2021classification}.
The RFF method consists of mapping data features to a finite-dimensional space where the inner product approximates a given kernel~\citep{rahimi2007rff}.
Such a map can be written as $\vb*{\varphi}_\text{rff}: \mathcal{X} \to \mathbb{R}^D$, where $D$ is some number of dimensions, such that $k(\vb*{x},\vb*{y}) \approx \vb*{\varphi}_\text{rff}(\vb*{x})\cdot\vb*{\varphi}_\text{rff}(\vb*{y})$, for some given shift-invariant kernel function $k: \mathcal{X}\times\mathcal{X}\to\mathbb{R}$.
This result is supported on Bochner's theorem~\citep{reed1975ii}, which affirms that a shift-invariant kernel $k$ is related to a particular probability measure $p(\vb*{w})$ through the Fourier transform.
This allows us to write the $i$-th component of $\vb*{\varphi}_\text{rff}$ as
\begin{align}
 \vb*{\varphi}_{\text{rff}\,i}(\vb*{x}) = \sqrt{\frac{2}{D}}\cos(\vb*{x}\cdot\vb*{w}_i + b_i),
\end{align}
where $\vb*{w}_i$ is sampled from $p$, and $b_i$ is sampled uniformly from $[0, 2\pi)$.
Finally, the RFFs obtained through $\vb*{\varphi}_{\text{rff}}$ can be used to define a QFM, for instance, through a binarised amplitude encoding.

In the case of the 1D data shown in~\cref{fig:circuits}(d), we can define the QFM $\psi(x)$ through
\begin{align}
    \psi(x) = \sum_{i=0}^{D-1} \vb*{\varphi}_{\text{rff}\,i+1}(x) \ket{\tilde{i}},
\end{align}
where $\ket{\tilde{i}}$ is the decimal representation of a bit string of length $\log_2 D$\footnote{In this work, $D=8$. 
Thus, $\ket{\tilde{0}} = \ket{0,0,0}, \ket{\tilde{1}} = \ket{0,0,1}, \ldots,\ket{\tilde{7}} = \ket{1,1,1}$. If $D$ is increased, more qubits will be needed, and the depth of the circuit will increase. This will be reflected in a much noisier estimation of the density.}. 
Remarkably, as we proved in~\citep{gonzalez2021learning}, this technique enables the approximation of any probability distribution using finite-dimensional density matrices at the core of the algorithm.

We chose the map $\vb*{\varphi}_\text{rff}$ to approximate the Gaussian kernel, with a given parameter $\gamma=80$, such that $\vb*{\varphi}_\text{rff}(x)\cdot\vb*{\varphi}_\text{rff}(x') \approx e^{-\gamma(x-x')^2}$~\citep{rahimi2008weighted}.
A total of eight RFFs were used so that the circuit in~\cref{fig:circuits}(c) consisted of three qubits.

In~\cref{fig:GaussianPreds} we show the density estimation carried out in three different ways. 
The three panels correspond to a classical simulation of the density estimation quantum circuit on the left, a classical simulation of the density estimation noisy quantum circuit on the middle (the noise model is the same as the one for IBM Bogotá, with differences in the probabilities and noisy quantum gates), and the density estimation carried out on the IBM Lima quantum device on the right.
In the three cases we get a good approximation of the probability density function from which training data was sampled.
The discrepancy between the kernel density estimation lines and the quantum circuit ones, even in the ideal case (right panel of~\cref{fig:GaussianPreds}), comes from approximating the Gaussian kernel with a small number of random Fourier features.
Finally, as in the classification case, we see that the noise model provided by IBM is far from simulating the actual behaviour of the quantum circuit.

\section{Discussion}\label{sec:discussion}

QFMs play a central role in this work, as they provide a solution to the problem of encoding classical data into quantum states of qubits.
Nonetheless, a calculation of the complete state is required prior to physically encoding the classical data into the quantum computer. 
This sole fact puts in danger the algorithmic advantage of our proposal running on a quantum computer versus running on a classical computer, due to the easy classical access to the wave function entries of the data set state $\ket{\Psi}$~\citep{cotler2021revisiting}.

As we mentioned before, the preparation of $\ket{\Psi}$ on a quantum computer can be done using several arbitrary quantum state preparation methods.
This is only done once.
If the data increases, so that a new state $\ket{\Psi_\text{new}}$ needs to be prepared, one can consider the simpler problem of preparing $\ket{\Psi_\text{new}}$ with $\ket{\Psi}$ as an initial state, instead of the usual initial state $\ket{\vb*{0}}$~\citep{haug2021optimal}.

The preparation of $\psi(\vb*{x})$ or $\phi(k)$ directly challenges the scalability of our proposal.
In this work, we have prepared $\phi(k)$ using one-hot encoding, which is a completely deterministic QFM with $O(1)$ gates.
However, the preparation of $\psi(\vb*{x})$ requires exponentially many quantum gates as a function of the number of qubits of the QFM's target physical system~\citep{shende2006synthesis}.
We used such arbitrary state preparation algorithms in the experiments of this work for illustration; however, this procedure is not scalable.
Instead, we can consider a parameterised quantum circuit $W(\vb*{\theta}, \vb*{x})$ that maps data $\vb*{x}$ and parameters $\vb*{\theta}$ in the angles of parameterised quantum gates.
Then, by minimising $\min_{\vb*{\theta}}\sum_i d(W(\vb*{\theta}, \vb*{x}), \psi(\vb*{x}))$, where $d(\bullet, \bullet)$ is a distance (fidelity~\citep{rakyta2022efficient}, KL divergence of the probability distributions represented by the states~\citep{liu2018differentiable}, classical shadows~\citep{li2021vsql,sack2022}, among others), one is able to obtain a variational circuit $W(\vb*{\theta}, \vb*{x})$ that acts as a primitive circuit to approximately apply QFMs to new data points $\vb*{x}_\star$ without investing exponential resources.

This proposed setup would use the primitives for preparing the data set state $\ket{\Psi}$, and for preparing the quantum state of a single data point $\psi(\vb*{x})$ to perform density estimation and classification, as shown in this paper. 
The numerical heavy-lifting that is exponential in the number of qubits of the system would need to be done just once when preparing the circuit primitives.
However, classifying or estimating the density of a new data sample would involve just the evaluation of the primitive circuits.
Of course, the feasibility of using this method for large scale quantum machine learning is subjected to the progress of training parameterised quantum circuits, which amounts to overcoming the barren plateau problem~\citep{sack2022,haug2021optimal,Sim2021adaptive,zhu2019training,Grant2019initialization,thanasilp2021}.

\section{Conclusions}\label{sec:conclusions}

Quantum machine learning has dominantly focused on making quantum versions of classical machine learning algorithms, most of which use gradient-based optimisation of parameters.
Recently, due to vanishing gradient issues, the community has started to switch to gradient-free techniques to address supervised and unsupervised learning with quantum hardware.
This work departs from the general idea of optimising parameters and exploits the intrinsic relation between geometry and probability that quantum theory offers.
For this, we implemented a method~\citep{gonzalez2021classification,gonzalez2021learning} to perform density estimation and data classification using quantum hardware.
This was achieved through the deterministic preparation of a quantum state that represents the information contained in a classical training data set and a quantum state that represents the information of a single point to be classified or whose probability density is to be estimated.
These quantum states are obtained by applying a quantum feature map to classical data points and are prepared using arbitrary quantum state algorithms.

One of the outstanding advantages of this method is the ability to approximate the probability distribution of arbitrarily large training data sets into finite-dimensional quantum states.
We demonstrated density estimation and classification with toy data sets using quantum circuits of three qubits.
We confirmed that the method's performance on real quantum devices suffered from decoherence, as expected.
However, the noise models provided by IBM's Qiskit are far from describing the actual behaviour of the quantum device for the applications we explored.
This shows that, even though the theory of open quantum systems has been well established, its practical application to large quantum systems has been a challenge.
Thus, our work adheres to the experimental evidence that more effective noise models are needed to simulate decoherence in quantum circuits.

Regarding possible quantum advantages, we acknowledge that the preparation of arbitrary quantum states can lead to the performance degradation of our method. 
Nonetheless, the exponential effort needed to prepare the quantum state of the training data set needs to be done only once.
Furthermore, we argued that the effort to prepare the quantum state of a new data point (to be classified or whose probability density is to be estimated) could also be made only just once by training a variational quantum circuit that performs the desired quantum feature map on an arbitrary input.
However, the feasibility of this alternative is subject to the advance of methods to train variational quantum circuits avoiding the barren plateau problem.

\backmatter



\section*{Statements and Declarations}

\begin{itemize}
    \item \textbf{Competing Interests:} The authors declare no competing interests.
\end{itemize}

\begin{appendices}

\section{One-hot Encoding Quantum Feature Map}\label{sec:OHEQFM}

Suppose that the classification problem considers $K$ classes (2 for the XOR data set in~\cref{fig:circuits}(b)).
Then, the one-hot encoding map is given by
\begin{align}
    \phi_\text{o.h.e.}(y_i) = \bigotimes_{j=1}^K(\delta_{j, y_i}\ket{1} + (1 - \delta_{j, y_i})\ket{0}),
\end{align}
where $\delta_{i,j} = 1$ if $i=j$ and is 0 otherwise.
The advantage of this QFM for class labels over other QFMs is that we no longer need to prepare the unitary $U_k$ to estimate the probability that a new data point $\vb*{x}_\star$ is of class $k$. Instead, let $\vb*{b}_k$ be the bit string defined as
\begin{align}
    \vb*{b}_k = \underbrace{0\ldots 0}_{N_\mathcal{X}}\underbrace{0\ldots0}_{k-1}1\underbrace{0\ldots0}_{N_\mathcal{Y}-k}.
\end{align}
Then, the estimated probability becomes
\begin{align}
    \hat{P}(k\vert \vb*{x}_\star) = M_{\vb*{b}_k}/M.
\end{align}





\end{appendices}

\bibliography{refs}


\begin{thebibliography}{57}
\ifx \bisbn   \undefined \def \bisbn  #1{ISBN #1}\fi
\ifx \binits  \undefined \def \binits#1{#1}\fi
\ifx \bauthor  \undefined \def \bauthor#1{#1}\fi
\ifx \batitle  \undefined \def \batitle#1{#1}\fi
\ifx \bjtitle  \undefined \def \bjtitle#1{#1}\fi
\ifx \bvolume  \undefined \def \bvolume#1{\textbf{#1}}\fi
\ifx \byear  \undefined \def \byear#1{#1}\fi
\ifx \bissue  \undefined \def \bissue#1{#1}\fi
\ifx \bfpage  \undefined \def \bfpage#1{#1}\fi
\ifx \blpage  \undefined \def \blpage #1{#1}\fi
\ifx \burl  \undefined \def \burl#1{\textsf{#1}}\fi
\ifx \doiurl  \undefined \def \doiurl#1{\url{https://doi.org/#1}}\fi
\ifx \betal  \undefined \def \betal{\textit{et al.}}\fi
\ifx \binstitute  \undefined \def \binstitute#1{#1}\fi
\ifx \binstitutionaled  \undefined \def \binstitutionaled#1{#1}\fi
\ifx \bctitle  \undefined \def \bctitle#1{#1}\fi
\ifx \beditor  \undefined \def \beditor#1{#1}\fi
\ifx \bpublisher  \undefined \def \bpublisher#1{#1}\fi
\ifx \bbtitle  \undefined \def \bbtitle#1{#1}\fi
\ifx \bedition  \undefined \def \bedition#1{#1}\fi
\ifx \bseriesno  \undefined \def \bseriesno#1{#1}\fi
\ifx \blocation  \undefined \def \blocation#1{#1}\fi
\ifx \bsertitle  \undefined \def \bsertitle#1{#1}\fi
\ifx \bsnm \undefined \def \bsnm#1{#1}\fi
\ifx \bsuffix \undefined \def \bsuffix#1{#1}\fi
\ifx \bparticle \undefined \def \bparticle#1{#1}\fi
\ifx \barticle \undefined \def \barticle#1{#1}\fi
\bibcommenthead
\ifx \bconfdate \undefined \def \bconfdate #1{#1}\fi
\ifx \botherref \undefined \def \botherref #1{#1}\fi
\ifx \url \undefined \def \url#1{\textsf{#1}}\fi
\ifx \bchapter \undefined \def \bchapter#1{#1}\fi
\ifx \bbook \undefined \def \bbook#1{#1}\fi
\ifx \bcomment \undefined \def \bcomment#1{#1}\fi
\ifx \oauthor \undefined \def \oauthor#1{#1}\fi
\ifx \citeauthoryear \undefined \def \citeauthoryear#1{#1}\fi
\ifx \endbibitem  \undefined \def \endbibitem {}\fi
\ifx \bconflocation  \undefined \def \bconflocation#1{#1}\fi
\ifx \arxivurl  \undefined \def \arxivurl#1{\textsf{#1}}\fi
\csname PreBibitemsHook\endcsname

\bibitem{bharti2022noisy}
\begin{barticle}
\bauthor{\bsnm{Bharti}, \binits{K.}},
\bauthor{\bsnm{Cervera-Lierta}, \binits{A.}},
\bauthor{\bsnm{Kyaw}, \binits{T.H.}},
\bauthor{\bsnm{Haug}, \binits{T.}},
\bauthor{\bsnm{Alperin-Lea}, \binits{S.}},
\bauthor{\bsnm{Anand}, \binits{A.}},
\bauthor{\bsnm{Degroote}, \binits{M.}},
\bauthor{\bsnm{Heimonen}, \binits{H.}},
\bauthor{\bsnm{Kottmann}, \binits{J.S.}},
\bauthor{\bsnm{Menke}, \binits{T.}},
\bauthor{\bsnm{Mok}, \binits{W.-K.}},
\bauthor{\bsnm{Sim}, \binits{S.}},
\bauthor{\bsnm{Kwek}, \binits{L.-C.}},
\bauthor{\bsnm{Aspuru-Guzik}, \binits{A.}}:
\batitle{Noisy intermediate-scale quantum algorithms}.
\bjtitle{Rev. Mod. Phys.}
\bvolume{94},
\bfpage{015004}
(\byear{2022}).
\doiurl{10.1103/RevModPhys.94.015004}
\end{barticle}
\endbibitem

\bibitem{huang2021information}
\begin{barticle}
\bauthor{\bsnm{Huang}, \binits{H.-Y.}},
\bauthor{\bsnm{Kueng}, \binits{R.}},
\bauthor{\bsnm{Preskill}, \binits{J.}}:
\batitle{Information-theoretic bounds on quantum advantage in machine
  learning}.
\bjtitle{Phys. Rev. Lett.}
\bvolume{126},
\bfpage{190505}
(\byear{2021}).
\doiurl{10.1103/PhysRevLett.126.190505}
\end{barticle}
\endbibitem

\bibitem{huang2021power}
\begin{barticle}
\bauthor{\bsnm{Huang}, \binits{H.-Y.}},
\bauthor{\bsnm{Broughton}, \binits{M.}},
\bauthor{\bsnm{Mohseni}, \binits{M.}},
\bauthor{\bsnm{Babbush}, \binits{R.}},
\bauthor{\bsnm{Boixo}, \binits{S.}},
\bauthor{\bsnm{Neven}, \binits{H.}},
\bauthor{\bsnm{McClean}, \binits{J.R.}}:
\batitle{Power of data in quantum machine learning}.
\bjtitle{Nature Communications}
\bvolume{12}(\bissue{1}),
\bfpage{2631}
(\byear{2021}).
\doiurl{10.1038/s41467-021-22539-9}
\end{barticle}
\endbibitem

\bibitem{huang2021quantum}
\begin{botherref}
\oauthor{\bsnm{Huang}, \binits{H.-Y.}},
\oauthor{\bsnm{Broughton}, \binits{M.}},
\oauthor{\bsnm{Cotler}, \binits{J.}},
\oauthor{\bsnm{Chen}, \binits{S.}},
\oauthor{\bsnm{Li}, \binits{J.}},
\oauthor{\bsnm{Mohseni}, \binits{M.}},
\oauthor{\bsnm{Neven}, \binits{H.}},
\oauthor{\bsnm{Babbush}, \binits{R.}},
\oauthor{\bsnm{Kueng}, \binits{R.}},
\oauthor{\bsnm{Preskill}, \binits{J.}},
\oauthor{\bsnm{McClean}, \binits{J.R.}}:
Quantum advantage in learning from experiments.
arXiv
(2021).
\doiurl{10.48550/arxiv.2112.00778}.
\url{https://arxiv.org/abs/2112.00778}
\end{botherref}
\endbibitem

\bibitem{rebentrost2014quantum}
\begin{barticle}
\bauthor{\bsnm{Rebentrost}, \binits{P.}},
\bauthor{\bsnm{Mohseni}, \binits{M.}},
\bauthor{\bsnm{Lloyd}, \binits{S.}}:
\batitle{Quantum support vector machine for big data classification}.
\bjtitle{Physical review letters}
\bvolume{113}(\bissue{13}),
\bfpage{130503}
(\byear{2014})
\end{barticle}
\endbibitem

\bibitem{schuld2015introduction}
\begin{barticle}
\bauthor{\bsnm{Schuld}, \binits{M.}},
\bauthor{\bsnm{Sinayskiy}, \binits{I.}},
\bauthor{\bsnm{Petruccione}, \binits{F.}}:
\batitle{An introduction to quantum machine learning}.
\bjtitle{Contemporary Physics}
\bvolume{56}(\bissue{2}),
\bfpage{172}--\blpage{185}
(\byear{2015})
{\href{https://arxiv.org/abs/https://doi.org/10.1080/00107514.2014.964942}{{https://doi.org/10.1080/00107514.2014.964942}}}.
\doiurl{10.1080/00107514.2014.964942}
\end{barticle}
\endbibitem

\bibitem{Dunjko2018machinelearning}
\begin{barticle}
\bauthor{\bsnm{Dunjko}, \binits{V.}},
\bauthor{\bsnm{Briegel}, \binits{H.J.}}:
\batitle{Machine learning {\&} artificial intelligence in the quantum domain: a
  review of recent progress}.
\bjtitle{Reports on Progress in Physics}
\bvolume{81}(\bissue{7}),
\bfpage{074001}
(\byear{2018}).
\doiurl{10.1088/1361-6633/aab406}
\end{barticle}
\endbibitem

\bibitem{biamonte2017quantum}
\begin{barticle}
\bauthor{\bsnm{Biamonte}, \binits{J.}},
\bauthor{\bsnm{Wittek}, \binits{P.}},
\bauthor{\bsnm{Pancotti}, \binits{N.}},
\bauthor{\bsnm{Rebentrost}, \binits{P.}},
\bauthor{\bsnm{Wiebe}, \binits{N.}},
\bauthor{\bsnm{Lloyd}, \binits{S.}}:
\batitle{Quantum machine learning}.
\bjtitle{Nature}
\bvolume{549}(\bissue{7671}),
\bfpage{195}--\blpage{202}
(\byear{2017})
\end{barticle}
\endbibitem

\bibitem{Benedetti2019parameterized}
\begin{botherref}
Parameterized quantum circuits as machine learning models.
Quantum Science and Technology
\textbf{4}(4),
043001
(2019).
\doiurl{10.1088/2058-9565/ab4eb5}
\end{botherref}
\endbibitem

\bibitem{caro2021generalization}
\begin{botherref}
\oauthor{\bsnm{Caro}, \binits{M.C.}},
\oauthor{\bsnm{Huang}, \binits{H.-Y.}},
\oauthor{\bsnm{Cerezo}, \binits{M.}},
\oauthor{\bsnm{Sharma}, \binits{K.}},
\oauthor{\bsnm{Sornborger}, \binits{A.}},
\oauthor{\bsnm{Cincio}, \binits{L.}},
\oauthor{\bsnm{Coles}, \binits{P.J.}}:
Generalization in quantum machine learning from few training data
(2021)
\end{botherref}
\endbibitem

\bibitem{mcclean2018barren}
\begin{barticle}
\bauthor{\bsnm{McClean}, \binits{J.R.}},
\bauthor{\bsnm{Boixo}, \binits{S.}},
\bauthor{\bsnm{Smelyanskiy}, \binits{V.N.}},
\bauthor{\bsnm{Babbush}, \binits{R.}},
\bauthor{\bsnm{Neven}, \binits{H.}}:
\batitle{Barren plateaus in quantum neural network training landscapes}.
\bjtitle{Nature communications}
\bvolume{9}(\bissue{1}),
\bfpage{1}--\blpage{6}
(\byear{2018})
\end{barticle}
\endbibitem

\bibitem{franken2020gradient}
\begin{botherref}
\oauthor{\bsnm{Franken}, \binits{L.}},
\oauthor{\bsnm{Georgiev}, \binits{B.}},
\oauthor{\bsnm{Muecke}, \binits{S.}},
\oauthor{\bsnm{Wolter}, \binits{M.}},
\oauthor{\bsnm{Piatkowski}, \binits{N.}},
\oauthor{\bsnm{Bauckhage}, \binits{C.}}:
Gradient-free quantum optimization on NISQ devices.
arXiv
(2020).
\doiurl{10.48550/arxiv.2012.13453}.
\url{https://arxiv.org/abs/2012.13453}
\end{botherref}
\endbibitem

\bibitem{benedetti2019generative}
\begin{barticle}
\bauthor{\bsnm{Benedetti}, \binits{M.}},
\bauthor{\bsnm{Garcia-Pintos}, \binits{D.}},
\bauthor{\bsnm{Perdomo}, \binits{O.}},
\bauthor{\bsnm{Leyton-Ortega}, \binits{V.}},
\bauthor{\bsnm{Nam}, \binits{Y.}},
\bauthor{\bsnm{Perdomo-Ortiz}, \binits{A.}}:
\batitle{A generative modeling approach for benchmarking and training shallow
  quantum circuits}.
\bjtitle{npj Quantum Information}
\bvolume{5}(\bissue{1}),
\bfpage{45}
(\byear{2019}).
\doiurl{10.1038/s41534-019-0157-8}
\end{barticle}
\endbibitem

\bibitem{peruzzo2014variational}
\begin{barticle}
\bauthor{\bsnm{Peruzzo}, \binits{A.}},
\bauthor{\bsnm{McClean}, \binits{J.}},
\bauthor{\bsnm{Shadbolt}, \binits{P.}},
\bauthor{\bsnm{Yung}, \binits{M.-H.}},
\bauthor{\bsnm{Zhou}, \binits{X.-Q.}},
\bauthor{\bsnm{Love}, \binits{P.J.}},
\bauthor{\bsnm{Aspuru-Guzik}, \binits{A.}},
\bauthor{\bsnm{O'Brien}, \binits{J.L.}}:
\batitle{A variational eigenvalue solver on a photonic quantum processor}.
\bjtitle{Nature Communications}
\bvolume{5}(\bissue{1}),
\bfpage{4213}
(\byear{2014}).
\doiurl{10.1038/ncomms5213}
\end{barticle}
\endbibitem

\bibitem{khatri2019quantum}
\begin{barticle}
\bauthor{\bsnm{Khatri}, \binits{S.}},
\bauthor{\bsnm{LaRose}, \binits{R.}},
\bauthor{\bsnm{Poremba}, \binits{A.}},
\bauthor{\bsnm{Cincio}, \binits{L.}},
\bauthor{\bsnm{Sornborger}, \binits{A.T.}},
\bauthor{\bsnm{Coles}, \binits{P.J.}}:
\batitle{Quantum-assisted quantum compiling}.
\bjtitle{Quantum}
\bvolume{3},
\bfpage{140}
(\byear{2019})
\end{barticle}
\endbibitem

\bibitem{leyton2021robust}
\begin{barticle}
\bauthor{\bsnm{Leyton-Ortega}, \binits{V.}},
\bauthor{\bsnm{Perdomo-Ortiz}, \binits{A.}},
\bauthor{\bsnm{Perdomo}, \binits{O.}}:
\batitle{Robust implementation of generative modeling with parametrized quantum
  circuits}.
\bjtitle{Quantum Machine Intelligence}
\bvolume{3}(\bissue{1}),
\bfpage{17}
(\byear{2021}).
\doiurl{10.1007/s42484-021-00040-2}
\end{barticle}
\endbibitem

\bibitem{Arrasmith2021effectofbarren}
\begin{barticle}
\bauthor{\bsnm{Arrasmith}, \binits{A.}},
\bauthor{\bsnm{Cerezo}, \binits{M.}},
\bauthor{\bsnm{Czarnik}, \binits{P.}},
\bauthor{\bsnm{Cincio}, \binits{L.}},
\bauthor{\bsnm{Coles}, \binits{P.J.}}:
\batitle{Effect of barren plateaus on gradient-free optimization}.
\bjtitle{{Quantum}}
\bvolume{5},
\bfpage{558}
(\byear{2021}).
\doiurl{10.22331/q-2021-10-05-558}
\end{barticle}
\endbibitem

\bibitem{marrero2020entanglement}
\begin{botherref}
\oauthor{\bsnm{Marrero}, \binits{C.O.}},
\oauthor{\bsnm{Kieferová}, \binits{M.}},
\oauthor{\bsnm{Wiebe}, \binits{N.}}:
Entanglement Induced Barren Plateaus.
arXiv
(2020).
\doiurl{10.48550/arxiv.2010.15968}.
\url{https://arxiv.org/abs/2010.15968}
\end{botherref}
\endbibitem

\bibitem{cerezo2021cost}
\begin{barticle}
\bauthor{\bsnm{Cerezo}, \binits{M.}},
\bauthor{\bsnm{Sone}, \binits{A.}},
\bauthor{\bsnm{Volkoff}, \binits{T.}},
\bauthor{\bsnm{Cincio}, \binits{L.}},
\bauthor{\bsnm{Coles}, \binits{P.J.}}:
\batitle{Cost function dependent barren plateaus in shallow parametrized
  quantum circuits}.
\bjtitle{Nature Communications}
\bvolume{12}(\bissue{1}),
\bfpage{1791}
(\byear{2021}).
\doiurl{10.1038/s41467-021-21728-w}
\end{barticle}
\endbibitem

\bibitem{gonzalez2021classification}
\begin{barticle}
\bauthor{\bsnm{Gonz{\'a}lez}, \binits{F.A.}},
\bauthor{\bsnm{Vargas-Calder{\'o}n}, \binits{V.}},
\bauthor{\bsnm{Vinck-Posada}, \binits{H.}}:
\batitle{Classification with quantum measurements}.
\bjtitle{Journal of the Physical Society of Japan}
\bvolume{90}(\bissue{4}),
\bfpage{044002}
(\byear{2021})
\end{barticle}
\endbibitem

\bibitem{gonzalez2021learning}
\begin{botherref}
\oauthor{\bsnm{Gonz{\'{a}}lez}, \binits{F.A.}},
\oauthor{\bsnm{Gallego}, \binits{A.}},
\oauthor{\bsnm{Toledo{-}Cort{\'{e}}s}, \binits{S.}},
\oauthor{\bsnm{Vargas{-}Calder{\'{o}}n}, \binits{V.}}:
Learning with density matrices and random features.
CoRR
\textbf{abs/2102.04394}
(2021)
{\href{https://arxiv.org/abs/2102.04394}{{arXiv:2102.04394}}}
\end{botherref}
\endbibitem

\bibitem{sergioli2019binary}
\begin{barticle}
\bauthor{\bsnm{Sergioli}, \binits{G.}},
\bauthor{\bsnm{Giuntini}, \binits{R.}},
\bauthor{\bsnm{Freytes}, \binits{H.}}:
\batitle{A new quantum approach to binary classification}.
\bjtitle{PLOS ONE}
\bvolume{14}(\bissue{5}),
\bfpage{1}--\blpage{14}
(\byear{2019}).
\doiurl{10.1371/journal.pone.0216224}
\end{barticle}
\endbibitem

\bibitem{mengoni209kernel}
\begin{barticle}
\bauthor{\bsnm{Mengoni}, \binits{R.}},
\bauthor{\bsnm{Di~Pierro}, \binits{A.}}:
\batitle{Kernel methods in quantum machine learning}.
\bjtitle{Quantum Machine Intelligence}
\bvolume{1}(\bissue{3}),
\bfpage{65}--\blpage{71}
(\byear{2019}).
\doiurl{10.1007/s42484-019-00007-4}
\end{barticle}
\endbibitem

\bibitem{schuld2019qml}
\begin{barticle}
\bauthor{\bsnm{Schuld}, \binits{M.}},
\bauthor{\bsnm{Killoran}, \binits{N.}}:
\batitle{Quantum machine learning in feature hilbert spaces}.
\bjtitle{Phys. Rev. Lett.}
\bvolume{122},
\bfpage{040504}
(\byear{2019}).
\doiurl{10.1103/PhysRevLett.122.040504}
\end{barticle}
\endbibitem

\bibitem{havlivcek2019supervised}
\begin{barticle}
\bauthor{\bsnm{Havl{\'\i}{\v{c}}ek}, \binits{V.}},
\bauthor{\bsnm{C{\'o}rcoles}, \binits{A.D.}},
\bauthor{\bsnm{Temme}, \binits{K.}},
\bauthor{\bsnm{Harrow}, \binits{A.W.}},
\bauthor{\bsnm{Kandala}, \binits{A.}},
\bauthor{\bsnm{Chow}, \binits{J.M.}},
\bauthor{\bsnm{Gambetta}, \binits{J.M.}}:
\batitle{Supervised learning with quantum-enhanced feature spaces}.
\bjtitle{Nature}
\bvolume{567}(\bissue{7747}),
\bfpage{209}--\blpage{212}
(\byear{2019})
\end{barticle}
\endbibitem

\bibitem{blank2020quantum}
\begin{barticle}
\bauthor{\bsnm{Blank}, \binits{C.}},
\bauthor{\bsnm{Park}, \binits{D.K.}},
\bauthor{\bsnm{Rhee}, \binits{J.-K.K.}},
\bauthor{\bsnm{Petruccione}, \binits{F.}}:
\batitle{Quantum classifier with tailored quantum kernel}.
\bjtitle{npj Quantum Information}
\bvolume{6}(\bissue{1}),
\bfpage{41}
(\byear{2020}).
\doiurl{10.1038/s41534-020-0272-6}
\end{barticle}
\endbibitem

\bibitem{parzen1962}
\begin{barticle}
\bauthor{\bsnm{Parzen}, \binits{E.}}:
\batitle{{On Estimation of a Probability Density Function and Mode}}.
\bjtitle{The Annals of Mathematical Statistics}
\bvolume{33}(\bissue{3}),
\bfpage{1065}--\blpage{1076}
(\byear{1962}).
\doiurl{10.1214/aoms/1177704472}
\end{barticle}
\endbibitem

\bibitem{rosenblatt1956}
\begin{barticle}
\bauthor{\bsnm{Rosenblatt}, \binits{M.}}:
\batitle{{Remarks on Some Nonparametric Estimates of a Density Function}}.
\bjtitle{The Annals of Mathematical Statistics}
\bvolume{27}(\bissue{3}),
\bfpage{832}--\blpage{837}
(\byear{1956}).
\doiurl{10.1214/aoms/1177728190}
\end{barticle}
\endbibitem

\bibitem{barenco1995qrdecomposition}
\begin{barticle}
\bauthor{\bsnm{Barenco}, \binits{A.}},
\bauthor{\bsnm{Bennett}, \binits{C.H.}},
\bauthor{\bsnm{Cleve}, \binits{R.}},
\bauthor{\bsnm{DiVincenzo}, \binits{D.P.}},
\bauthor{\bsnm{Margolus}, \binits{N.}},
\bauthor{\bsnm{Shor}, \binits{P.}},
\bauthor{\bsnm{Sleator}, \binits{T.}},
\bauthor{\bsnm{Smolin}, \binits{J.A.}},
\bauthor{\bsnm{Weinfurter}, \binits{H.}}:
\batitle{Elementary gates for quantum computation}.
\bjtitle{Phys. Rev. A}
\bvolume{52},
\bfpage{3457}--\blpage{3467}
(\byear{1995}).
\doiurl{10.1103/PhysRevA.52.3457}
\end{barticle}
\endbibitem

\bibitem{mottonen2004sincosDecomposition}
\begin{barticle}
\bauthor{\bsnm{M\"ott\"onen}, \binits{M.}},
\bauthor{\bsnm{Vartiainen}, \binits{J.J.}},
\bauthor{\bsnm{Bergholm}, \binits{V.}},
\bauthor{\bsnm{Salomaa}, \binits{M.M.}}:
\batitle{Quantum circuits for general multiqubit gates}.
\bjtitle{Phys. Rev. Lett.}
\bvolume{93},
\bfpage{130502}
(\byear{2004}).
\doiurl{10.1103/PhysRevLett.93.130502}
\end{barticle}
\endbibitem

\bibitem{krol2022efficient}
\begin{botherref}
\oauthor{\bsnm{Krol}, \binits{A.M.}},
\oauthor{\bsnm{Sarkar}, \binits{A.}},
\oauthor{\bsnm{Ashraf}, \binits{I.}},
\oauthor{\bsnm{Al-Ars}, \binits{Z.}},
\oauthor{\bsnm{Bertels}, \binits{K.}}:
Efficient decomposition of unitary matrices in quantum circuit compilers.
Applied Sciences
\textbf{12}(2)
(2022).
\doiurl{10.3390/app12020759}
\end{botherref}
\endbibitem

\bibitem{li2013decomposition}
\begin{barticle}
\bauthor{\bsnm{Li}, \binits{C.-K.}},
\bauthor{\bsnm{Roberts}, \binits{R.}},
\bauthor{\bsnm{Yin}, \binits{X.}}:
\batitle{Decomposition of unitary matrices and quantum gates}.
\bjtitle{International Journal of Quantum Information}
\bvolume{11}(\bissue{01}),
\bfpage{1350015}
(\byear{2013})
{\href{https://arxiv.org/abs/https://doi.org/10.1142/S0219749913500159}{{https://doi.org/10.1142/S0219749913500159}}}.
\doiurl{10.1142/S0219749913500159}
\end{barticle}
\endbibitem

\bibitem{shende2006synthesis}
\begin{barticle}
\bauthor{\bsnm{Shende}, \binits{V.V.}},
\bauthor{\bsnm{Bullock}, \binits{S.S.}},
\bauthor{\bsnm{Markov}, \binits{I.L.}}:
\batitle{Synthesis of quantum-logic circuits}.
\bjtitle{IEEE Transactions on Computer-Aided Design of Integrated Circuits and
  Systems}
\bvolume{25}(\bissue{6}),
\bfpage{1000}--\blpage{1010}
(\byear{2006})
\end{barticle}
\endbibitem

\bibitem{Qiskit}
\begin{botherref}
\oauthor{\bsnm{Treinish}, \binits{M.}},
\oauthor{\bsnm{Gambetta}, \binits{J.}},
\oauthor{\bsnm{Nation}, \binits{P.}},
\oauthor{\bsnm{Kassebaum}, \binits{P.}},
\oauthor{\bsnm{qiskit-bot}},
\oauthor{\bsnm{Rodríguez}, \binits{D.M.}},
\oauthor{\bparticle{de~la} \bsnm{Puente~González}, \binits{S.}},
\oauthor{\bsnm{Hu}, \binits{S.}},
\oauthor{\bsnm{Krsulich}, \binits{K.}},
\oauthor{\bsnm{Zdanski}, \binits{L.}},
\oauthor{\bsnm{Yu}, \binits{J.}},
\oauthor{\bsnm{Garrison}, \binits{J.}},
\oauthor{\bsnm{Gacon}, \binits{J.}},
\oauthor{\bsnm{McKay}, \binits{D.}},
\oauthor{\bsnm{Gomez}, \binits{J.}},
\oauthor{\bsnm{Capelluto}, \binits{L.}},
\oauthor{\bsnm{Travis-S-IBM}},
\oauthor{\bsnm{Marques}, \binits{M.}},
\oauthor{\bsnm{Panigrahi}, \binits{A.}},
\oauthor{\bsnm{Lishman}, \binits{J.}},
\oauthor{\bsnm{lerongil}},
\oauthor{\bsnm{Rahman}, \binits{R.I.}},
\oauthor{\bsnm{Wood}, \binits{S.}},
\oauthor{\bsnm{Bello}, \binits{L.}},
\oauthor{\bsnm{Singh}, \binits{D.}},
\oauthor{\bsnm{Drew}},
\oauthor{\bsnm{Arbel}, \binits{E.}},
\oauthor{\bsnm{Schwarm}, \binits{J.}},
\oauthor{\bsnm{Daniel}, \binits{J.}},
\oauthor{\bsnm{George}, \binits{M.}}:
Qiskit/qiskit: Qiskit 0.34.2.
\doiurl{10.5281/zenodo.6027041}
\end{botherref}
\endbibitem

\bibitem{bausch2020fast}
\begin{botherref}
\oauthor{\bsnm{Bausch}, \binits{J.}}:
Fast Black-Box Quantum State Preparation.
arXiv
(2020).
\doiurl{10.48550/ARXIV.2009.10709}.
\url{https://arxiv.org/abs/2009.10709}
\end{botherref}
\endbibitem

\bibitem{araujo2021qsp}
\begin{barticle}
\bauthor{\bsnm{Araujo}, \binits{I.F.}},
\bauthor{\bsnm{Park}, \binits{D.K.}},
\bauthor{\bsnm{Petruccione}, \binits{F.}},
\bauthor{\bparticle{da} \bsnm{Silva}, \binits{A.J.}}:
\batitle{A divide-and-conquer algorithm for quantum state preparation}.
\bjtitle{Scientific Reports}
\bvolume{11}(\bissue{1}),
\bfpage{6329}
(\byear{2021}).
\doiurl{10.1038/s41598-021-85474-1}
\end{barticle}
\endbibitem

\bibitem{zhang2021lowdepth}
\begin{barticle}
\bauthor{\bsnm{Zhang}, \binits{X.-M.}},
\bauthor{\bsnm{Yung}, \binits{M.-H.}},
\bauthor{\bsnm{Yuan}, \binits{X.}}:
\batitle{Low-depth quantum state preparation}.
\bjtitle{Phys. Rev. Research}
\bvolume{3},
\bfpage{043200}
(\byear{2021}).
\doiurl{10.1103/PhysRevResearch.3.043200}
\end{barticle}
\endbibitem

\bibitem{Haug2020classifying}
\begin{barticle}
\bauthor{\bsnm{Haug}, \binits{T.}},
\bauthor{\bsnm{Mok}, \binits{W.-K.}},
\bauthor{\bsnm{You}, \binits{J.-B.}},
\bauthor{\bsnm{Zhang}, \binits{W.}},
\bauthor{\bsnm{Png}, \binits{C.E.}},
\bauthor{\bsnm{Kwek}, \binits{L.-C.}}:
\batitle{Classifying global state preparation via deep reinforcement learning}.
\bjtitle{Machine Learning: Science and Technology}
\bvolume{2}(\bissue{1}),
\bfpage{01}--\blpage{02}
(\byear{2020}).
\doiurl{10.1088/2632-2153/abc81f}
\end{barticle}
\endbibitem

\bibitem{rakyta2022efficient}
\begin{botherref}
\oauthor{\bsnm{Rakyta}, \binits{P.}},
\oauthor{\bsnm{Zimborás}, \binits{Z.}}:
Efficient quantum gate decomposition via adaptive circuit compression.
arXiv
(2022).
\doiurl{10.48550/arxiv.2203.04426}.
\url{https://arxiv.org/abs/2203.04426}
\end{botherref}
\endbibitem

\bibitem{shirakawa2021automatic}
\begin{botherref}
\oauthor{\bsnm{Shirakawa}, \binits{T.}},
\oauthor{\bsnm{Ueda}, \binits{H.}},
\oauthor{\bsnm{Yunoki}, \binits{S.}}:
Automatic quantum circuit encoding of a given arbitrary quantum state.
arXiv preprint arXiv:2112.14524
(2021)
\end{botherref}
\endbibitem

\bibitem{bogdanov2013}
\begin{bchapter}
\bauthor{\bsnm{Bogdanov}, \binits{Y.I.}},
\bauthor{\bsnm{Chernyavskiy}, \binits{A.Y.}},
\bauthor{\bsnm{Holevo}, \binits{A.}},
\bauthor{\bsnm{Lukichev}, \binits{V.F.}},
\bauthor{\bsnm{Orlikovsky}, \binits{A.A.}}:
\bctitle{{Modeling of quantum noise and the quality of hardware components of
  quantum computers}}.
In: \beditor{\bsnm{Orlikovsky}, \binits{A.A.}} (ed.)
\bbtitle{International Conference Micro- and Nano-Electronics 2012},
vol. \bseriesno{8700},
pp. \bfpage{404}--\blpage{415}.
\bpublisher{SPIE}, \blocation{???}
(\byear{2013}).
\doiurl{10.1117/12.2017414}.
\bcomment{International Society for Optics and Photonics}.
\burl{https://doi.org/10.1117/12.2017414}
\end{bchapter}
\endbibitem

\bibitem{dalibard1992}
\begin{barticle}
\bauthor{\bsnm{Dalibard}, \binits{J.}},
\bauthor{\bsnm{Castin}, \binits{Y.}},
\bauthor{\bsnm{M\o{}lmer}, \binits{K.}}:
\batitle{Wave-function approach to dissipative processes in quantum optics}.
\bjtitle{Phys. Rev. Lett.}
\bvolume{68},
\bfpage{580}--\blpage{583}
(\byear{1992}).
\doiurl{10.1103/PhysRevLett.68.580}
\end{barticle}
\endbibitem

\bibitem{molmer1993monte}
\begin{barticle}
\bauthor{\bsnm{M{\o}lmer}, \binits{K.}},
\bauthor{\bsnm{Castin}, \binits{Y.}},
\bauthor{\bsnm{Dalibard}, \binits{J.}}:
\batitle{Monte carlo wave-function method in quantum optics}.
\bjtitle{JOSA B}
\bvolume{10}(\bissue{3}),
\bfpage{524}--\blpage{538}
(\byear{1993})
\end{barticle}
\endbibitem

\bibitem{chow2015}
\begin{bchapter}
\bauthor{\bsnm{Chow}, \binits{J.M.}},
\bauthor{\bsnm{Srinivasan}, \binits{S.J.}},
\bauthor{\bsnm{Magesan}, \binits{E.}},
\bauthor{\bsnm{C{\'o}rcoles}, \binits{A.D.}},
\bauthor{\bsnm{Abraham}, \binits{D.W.}},
\bauthor{\bsnm{Gambetta}, \binits{J.M.}},
\bauthor{\bsnm{Steffen}, \binits{M.}}:
\bctitle{{Characterizing a four-qubit planar lattice for arbitrary error
  detection}}.
In: \beditor{\bsnm{Donkor}, \binits{E.}},
\beditor{\bsnm{Pirich}, \binits{A.R.}},
\beditor{\bsnm{Hayduk}, \binits{M.}} (eds.)
\bbtitle{Quantum Information and Computation XIII},
vol. \bseriesno{9500},
pp. \bfpage{315}--\blpage{323}.
\bpublisher{SPIE}, \blocation{???}
(\byear{2015}).
\doiurl{10.1117/12.2192740}.
\bcomment{International Society for Optics and Photonics}.
\burl{https://doi.org/10.1117/12.2192740}
\end{bchapter}
\endbibitem

\bibitem{berg2022probabilistic}
\begin{botherref}
\oauthor{\bsnm{Berg}, \binits{E.v.d.}},
\oauthor{\bsnm{Minev}, \binits{Z.K.}},
\oauthor{\bsnm{Kandala}, \binits{A.}},
\oauthor{\bsnm{Temme}, \binits{K.}}:
Probabilistic error cancellation with sparse pauli-lindblad models on noisy
  quantum processors.
arXiv preprint arXiv:2201.09866
(2022)
\end{botherref}
\endbibitem

\bibitem{rahimi2007rff}
\begin{bchapter}
\bauthor{\bsnm{Rahimi}, \binits{A.}},
\bauthor{\bsnm{Recht}, \binits{B.}}:
\bctitle{Random features for large-scale kernel machines}.
In: \bbtitle{Advances in Neural Information Processing Systems},
vol. \bseriesno{20},
pp. \bfpage{1160}--\blpage{1167}
(\byear{2007})
\end{bchapter}
\endbibitem

\bibitem{reed1975ii}
\begin{bbook}
\bauthor{\bsnm{Reed}, \binits{M.}},
\bauthor{\bsnm{Simon}, \binits{B.}}:
\bbtitle{II: Fourier Analysis, Self-Adjointness}
vol. \bseriesno{2},
(\byear{1975})
\end{bbook}
\endbibitem

\bibitem{rahimi2008weighted}
\begin{bchapter}
\bauthor{\bsnm{Rahimi}, \binits{A.}},
\bauthor{\bsnm{Recht}, \binits{B.}}:
\bctitle{Weighted sums of random kitchen sinks: Replacing minimization with
  randomization in learning}.
In: \beditor{\bsnm{Koller}, \binits{D.}},
\beditor{\bsnm{Schuurmans}, \binits{D.}},
\beditor{\bsnm{Bengio}, \binits{Y.}},
\beditor{\bsnm{Bottou}, \binits{L.}} (eds.)
\bbtitle{Advances in Neural Information Processing Systems},
vol. \bseriesno{21},
pp. \bfpage{1316}--\blpage{1323}
(\byear{2008})
\end{bchapter}
\endbibitem

\bibitem{cotler2021revisiting}
\begin{botherref}
\oauthor{\bsnm{Cotler}, \binits{J.}},
\oauthor{\bsnm{Huang}, \binits{H.-Y.}},
\oauthor{\bsnm{McClean}, \binits{J.R.}}:
Revisiting dequantization and quantum advantage in learning tasks.
arXiv preprint arXiv:2112.00811
(2021)
\end{botherref}
\endbibitem

\bibitem{haug2021optimal}
\begin{botherref}
\oauthor{\bsnm{Haug}, \binits{T.}},
\oauthor{\bsnm{Kim}, \binits{M.S.}}:
Optimal training of variational quantum algorithms without barren plateaus.
arXiv
(2021).
\doiurl{10.48550/arxiv.2104.14543}.
\url{https://arxiv.org/abs/2104.14543}
\end{botherref}
\endbibitem

\bibitem{liu2018differentiable}
\begin{barticle}
\bauthor{\bsnm{Liu}, \binits{J.-G.}},
\bauthor{\bsnm{Wang}, \binits{L.}}:
\batitle{Differentiable learning of quantum circuit born machines}.
\bjtitle{Phys. Rev. A}
\bvolume{98},
\bfpage{062324}
(\byear{2018}).
\doiurl{10.1103/PhysRevA.98.062324}
\end{barticle}
\endbibitem

\bibitem{li2021vsql}
\begin{bchapter}
\bauthor{\bsnm{Li}, \binits{G.}},
\bauthor{\bsnm{Song}, \binits{Z.}},
\bauthor{\bsnm{Wang}, \binits{X.}}:
\bctitle{Vsql: variational shadow quantum learning for classification}.
In: \bbtitle{Proceedings of the AAAI Conference on Artificial Intelligence},
vol. \bseriesno{35},
pp. \bfpage{8357}--\blpage{8365}
(\byear{2021})
\end{bchapter}
\endbibitem

\bibitem{sack2022}
\begin{botherref}
\oauthor{\bsnm{Sack}, \binits{S.H.}},
\oauthor{\bsnm{Medina}, \binits{R.A.}},
\oauthor{\bsnm{Michailidis}, \binits{A.A.}},
\oauthor{\bsnm{Kueng}, \binits{R.}},
\oauthor{\bsnm{Serbyn}, \binits{M.}}:
Avoiding barren plateaus using classical shadows.
arXiv
(2022).
\doiurl{10.48550/arxiv.2201.08194}.
\url{https://arxiv.org/abs/2201.08194}
\end{botherref}
\endbibitem

\bibitem{Sim2021adaptive}
\begin{barticle}
\bauthor{\bsnm{Sim}, \binits{S.}},
\bauthor{\bsnm{Romero}, \binits{J.}},
\bauthor{\bsnm{Gonthier}, \binits{J.F.}},
\bauthor{\bsnm{Kunitsa}, \binits{A.A.}}:
\batitle{Adaptive pruning-based optimization of parameterized quantum
  circuits}.
\bjtitle{Quantum Science and Technology}
\bvolume{6}(\bissue{2}),
\bfpage{025019}
(\byear{2021}).
\doiurl{10.1088/2058-9565/abe107}
\end{barticle}
\endbibitem

\bibitem{zhu2019training}
\begin{barticle}
\bauthor{\bsnm{Zhu}, \binits{D.}},
\bauthor{\bsnm{Linke}, \binits{N.M.}},
\bauthor{\bsnm{Benedetti}, \binits{M.}},
\bauthor{\bsnm{Landsman}, \binits{K.A.}},
\bauthor{\bsnm{Nguyen}, \binits{N.H.}},
\bauthor{\bsnm{Alderete}, \binits{C.H.}},
\bauthor{\bsnm{Perdomo-Ortiz}, \binits{A.}},
\bauthor{\bsnm{Korda}, \binits{N.}},
\bauthor{\bsnm{Garfoot}, \binits{A.}},
\bauthor{\bsnm{Brecque}, \binits{C.}},
\bauthor{\bsnm{Egan}, \binits{L.}},
\bauthor{\bsnm{Perdomo}, \binits{O.}},
\bauthor{\bsnm{Monroe}, \binits{C.}}:
\batitle{Training of quantum circuits on a hybrid quantum computer}.
\bjtitle{Science Advances}
\bvolume{5}(\bissue{10}),
\bfpage{9918}
(\byear{2019})
{\href{https://arxiv.org/abs/https://www.science.org/doi/pdf/10.1126/sciadv.aaw9918}{{https://www.science.org/doi/pdf/10.1126/sciadv.aaw9918}}}.
\doiurl{10.1126/sciadv.aaw9918}
\end{barticle}
\endbibitem

\bibitem{Grant2019initialization}
\begin{barticle}
\bauthor{\bsnm{Grant}, \binits{E.}},
\bauthor{\bsnm{Wossnig}, \binits{L.}},
\bauthor{\bsnm{Ostaszewski}, \binits{M.}},
\bauthor{\bsnm{Benedetti}, \binits{M.}}:
\batitle{An initialization strategy for addressing barren plateaus in
  parametrized quantum circuits}.
\bjtitle{{Quantum}}
\bvolume{3},
\bfpage{214}
(\byear{2019}).
\doiurl{10.22331/q-2019-12-09-214}
\end{barticle}
\endbibitem

\bibitem{thanasilp2021}
\begin{botherref}
\oauthor{\bsnm{Thanasilp}, \binits{S.}},
\oauthor{\bsnm{Wang}, \binits{S.}},
\oauthor{\bsnm{Nghiem}, \binits{N.A.}},
\oauthor{\bsnm{Coles}, \binits{P.J.}},
\oauthor{\bsnm{Cerezo}, \binits{M.}}:
Subtleties in the trainability of quantum machine learning models.
arXiv
(2021).
\doiurl{10.48550/ARXIV.2110.14753}.
\url{https://arxiv.org/abs/2110.14753}
\end{botherref}
\endbibitem

\end{thebibliography}


\end{document}